\documentclass[aip,reprint]{revtex4-1}

\draft 

\usepackage[T1]{fontenc}
\usepackage[latin9]{inputenc}
\usepackage{amsmath}
\usepackage{graphicx}
\usepackage{amssymb}
\usepackage{color,soul}
\usepackage{hyperref}

\begin{document}


\title{Theory of heterogeneous circuits with stochastic memristive devices} 



\author{Valeriy A. Slipko}
\email[]{vslipko@uni.opole.pl} \affiliation{Institute of Physics, Opole University, Opole 45-052, Poland}

\author{Yuriy V. Pershin}
\email[]{pershin@physics.sc.edu} \affiliation{Department of Physics and Astronomy, University of South Carolina, Columbia, SC 29208 USA}



\date{\today}

\begin{abstract}
We introduce an approach based on the Chapman-Kolmogorov equation to model heterogeneous stochastic circuits, namely, the circuits combining binary or multi-state stochastic memristive devices and continuum reactive components (capacitors and/or inductors). Such circuits are described in terms of occupation probabilities of memristive states that are  functions of reactive variables. As an illustrative example, the series circuit of a binary memristor and capacitor is considered in detail. Some analytical solutions are found. Our work offers a novel analytical/numerical tool for modeling complex stochastic networks, which may find a broad range of applications.
\end{abstract}


\maketitle 


\section{Introduction}

The well-known cycle-to-cycle variability of memristive devices is one of the major
obstacles for their use in non-volatile memories and other potential applications. While, traditionally, the memristive devices (memristors) and their
circuits have been mostly described in terms of deterministic models~\cite{chua76a,Kvatinsky15a,pershin09b,Strachan13a}, there is a growing evidence that at least in a
certain large group of memristive devices (such as the electrochemical metallization (ECM) cells~\cite{valov2011electrochemical}) the resistance switching
occurs stochastically, namely, at a random time interval from the pulse edge~\cite{jo2009programmable,gaba2013stochastic,gaba2014memristive}. Theoretically, the randomness in the resistance switching was implemented
in several models~\cite{menzel2014statistical,7827976,dowling2020probabilistic,dowling2020SPICE,Ntinas2020}.

Mathematically, stochastic processes can be  described by introducing a noise term~\cite{vanCampen} into the equations of memristor dynamics. Usually these noise terms correspond to a white or, in a more complicated case, colored noise. In the case of linear equations
(with respect to the internal state variables), the noisy dynamics  can be effectively
analysed  or even   solved analytically. For non-linear stochastic differential
equations, mainly numerical solutions can be found. This complicates tremendously the analysis
of general properties of processes that such equations describe.
In the cases when the stochastic processes are markovian, the evolution of probability distributions can be described by
a master equation, which was first applied to stochastic memristive networks by the present authors together with V.~Dowling~\cite{dowling2020probabilistic}.
However, the application range of such approach~\cite{dowling2020probabilistic} is limited to circuits composed of binary
or multi-state stochastic memristors, current and/or voltage sources, and non-linear components such as diodes.

 The purpose of the present paper is to generalize the method of master equation~\cite{dowling2020probabilistic} to more complex circuits that include also capacitors and/or indictors.
 Our main goal is to understand the behavior of complex stochastic circuits on average, as each particular realization of the circuit dynamics is not representative of the circuit behavior overall (as it depends on the particular realization of switching probabilities). The difficulty
stems from the fact that in any particular realization of circuit dynamics, the switching of memristive components depends on the values of
reactive variables, and vice-versa. Therefore, the statistical description is also necessary for the description of reactive variables. For this purpose, we utilize a set of  probability distribution functions.

This paper is organized in the following sections. In Sec.~\ref{sec:2}, the model of binary and multi-state stochastic memristors is presented. Section~\ref{sec:3} is the main part of this work, where the
evolution equations for a binary memristor-capacitor circuit are derived from the
Chapman-Kolmogorov equation (Subsec.~\ref{sec:31}), and their analytical solutions are found (Subsec.~\ref{sec:32}). The application of Sec.~\ref{sec:3} approach to more complex circuits is discussed in Sec.~\ref{sec:4} that concludes.

\section{Stochastic memristor model}\label{sec:2}

In the present paper we consider discrete stochastic memristors, whose resistance is characterized by $G$ values $R_i$, $i=0,..,G-1$. The transitions between two states, $i$ and $j$, are described in terms of the voltage-dependent transition rates, $\gamma_{i\rightarrow j}(V_M)\equiv\gamma_{ij}(V_M)$, where $V_M$ is the voltage across the memristor. Only
transitions between the adjacent states are allowed (e. g., $0\rightarrow 1$, $1\rightarrow 2$, ..., at $V_M>0$, etc.). The transition from one boundary state to another occurs sequentially through all intermediate states.

The binary stochastic memristors~\cite{dowling2020probabilistic} are the simplest case of discrete stochastic memristors.
Experiments have shown~\cite{jo2009programmable,gaba2013stochastic,gaba2014memristive} that in some electrochemical metallization cells the switching can be described by a probabilistic model with voltage-dependent switching rates
 \begin{eqnarray}
 \gamma_{0\rightarrow 1}(V_M)=\left\{ \begin{array}{cl}
\left( \tau_0 e^{-V_M/V_0}\right)^{-1},& V_M>0 \\
0 & \textnormal{otherwise}
\end{array}\right. \; , \label{eq:gamma01}\\
 \gamma_{1\rightarrow 0}(V_M)=\left\{ \begin{array}{cl}
\left( \tau_1 e^{-|V_M|/V_1}\right)^{-1},& V_M<0 \\
0 & \textnormal{otherwise}
\end{array}\right. \; . \label{eq:gamma10}
 \end{eqnarray}
Here, $\tau_{0(1)}$ and
$V_{0(1)}$ are constants. The interpretation of the above equations is  following: The probability to switch from $R_{off}\equiv R_0$ (state 0) to $R_{on}\equiv R_1$  (state 1) within the infinitesimal  time interval $\Delta t$ is  $\gamma_{0\rightarrow 1}(V_M) \Delta t$. The probability to switch in the opposite direction is defined similarly. Equations similar to Eqs. (\ref{eq:gamma01})-(\ref{eq:gamma10}) can be used for multi-state ($G>2$) stochastic memristors~\cite{dowling2020SPICE,Ntinas2020}.

\section{Binary Memristor-Capacitor Circuit} \label{sec:3}

In this Section we consider the circuit shown in Fig.~\ref{fig:1}(a), where M is a stochastic binary memristor (described by Eqs. (\ref{eq:gamma01}) and (\ref{eq:gamma10})),
and C is a linear capacitor.
It is assumed that the circuit is driven by a time-dependent voltage $V(t)$.

\subsection{Dynamical equations} \label{sec:31}

The circuit description is based on the probability distribution functions,
$p_{i}(q,t)$, $i=0,1$. The meaning of these functions is that
$p_{i}(q,t)\Delta q$ is the probability
to find the memristor in state $R_M=R_{i}$,
and capacitor charge $q$ in the interval from $q$ to $q+\Delta q$  at time $t$.
For the full probabilistic  description of memristive switching
it is necessary to introduce the transition probabilities
$P(R_{k},q_2,t_2|R_{j},q_1,t_1)\equiv P_{kj}(q_2,t_2|q_1,t_1) $ from the
memristor state $R_{j}$ and capacitor charge $q_1$ at time $t_1$ to
the memristor state $R_{k}$ and capacitor charge $q_2$ at time $t_2$.

The Chapman-Kolmogorov  equation~\cite{vanCampen} describing the Markov
process of stochastic switching can be written as
\begin{eqnarray}
p_{k}(q,t+\tau)=\sum\limits_{j=0}^{M-1} \int_{-\infty}^{+\infty} P_{kj}(q,t+\tau|r,t)p_{j}(r,t)\textnormal{d}r. \;
 \label{meq}
\end{eqnarray}

In order to satisfy the normalization condition
at any time $t$
\begin{eqnarray}
\sum\limits_{k=0}^{M-1} \int_{-\infty}^{+\infty} p_{k}(q,t)\textnormal{d}q=1,
 \label{norm}
\end{eqnarray}
 the transition probabilities
must satisfy for any $r$, $j$, $t$, and $\tau>0$ the corresponding conditions
\begin{eqnarray}
\sum\limits_{k=0}^{M-1}\int_{-\infty}^{+\infty} P_{kj}(q,t+\tau|r,t)\textnormal{d}q=1.
 \label{normtran}
\end{eqnarray}

It is easy to find the transition probabilities
for infinitesimally small $\tau>0$. There are four transition probabilities as the memristive system is binary.
For example, the transitional probability for switching from $R_{0}$ to $R_{1}$ is
equal to
  \begin{eqnarray}
P_{10}(q, t+\tau|r,t)=\gamma_{0\rightarrow 1}\left(V(t)-\frac{r}{C} \right)\tau \nonumber\\
\times\delta\left(q-r-\frac{\tau}{R_0}\left(V(t)-\frac{r}{C}\right)\right),
\tau\rightarrow +0,
\label{p21}
\end{eqnarray}
where $V(t)-r/C$ is the voltage drop across the memristor,  and the Dirac delta
function represents the change in the capacitor charge that is defined by Kirchhoff's law
\begin{eqnarray}
I\approx\frac{q-r}{\tau}=\frac{1}{R_0}\left(V(t)-\frac{r}{C}\right), \tau\rightarrow +
0.
 \label{eq13}
\end{eqnarray}

\begin{figure}[t]
\begin{center}
(a)\includegraphics[width=.15\columnwidth]{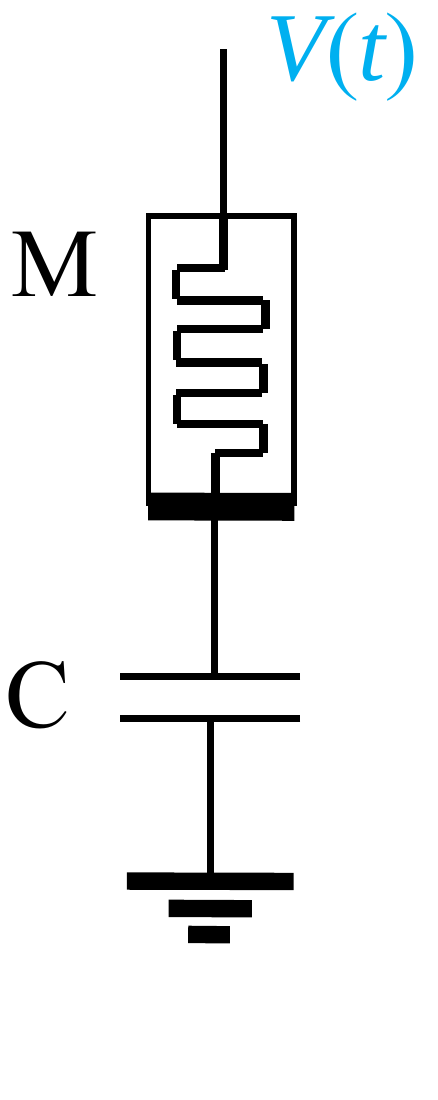}\;\;
(b)\includegraphics[width=.60\columnwidth]{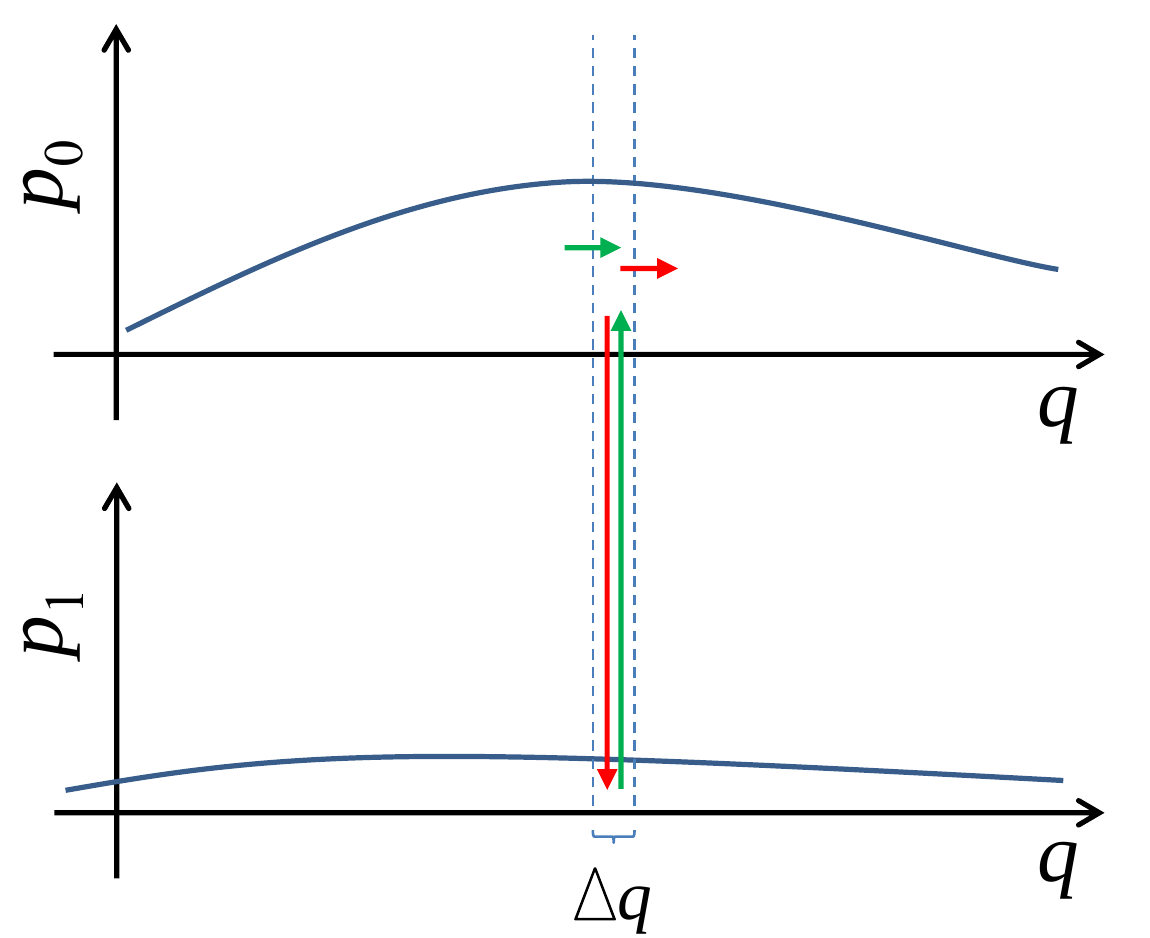}
\caption{\label{fig:1} (a) Schematics of the binary memristor-capacitor circuit. (b) Evolution scheme for $p_0(q,t)\Delta q$.
The interval $\Delta q$ is defined by two vertical lines. The arrows represent the flow of probability density.}
\end{center}
\end{figure}

Then, for the transitional probability from  $R_{0}$ to $R_{0}$ (no
switching) we can write (similarly to Eq. (\ref{p21}))
\begin{eqnarray}
P_{00}(q, t+\tau|r,t)=\left[1-\gamma_{0\rightarrow 1}\left(V(t)-\frac{r}{C}\right)\tau\right] \nonumber\\
\times\delta\left(q-r-\frac{\tau}{R_0}\left(V-\frac{r}{C}\right)\right),
\tau\rightarrow +0.
\label{p11}
\end{eqnarray}
The transitional probabilities $P_{12}$ and $P_{22}$ are obtained
by replacing $0\rightarrow 1$ and $1\rightarrow 0$ in Eqs. (\ref{p21}) and (\ref{p11}).
Note  that such expressions for the transitional probabilities
satisfy normalization conditions  (\ref{normtran}). Also we should note that
Eqs. (\ref{p21}) and (\ref{p11}) are valid only  up to the first order of magnitude of time  $\tau$, i.e. we should omit $\tau^2$-terms or higher while using
these equations. 

By substituting Eqs. (\ref{p21}) and (\ref{p11}) into the Chapman-Kolmogorov  equation (\ref{meq}) with $k=0$, and expanding it with respect to $\tau$ up to the first order in magnitude, we get the following
partial differential equation for the probability distribution function $p_{0}(q,t)$
\begin{eqnarray}
\frac{\partial p_0(q,t)}{\partial t}+\frac{\partial}{\partial q} \left( \frac{V_M}{R_{0}}p_0(q,t)\right) ~~~~~~~~~~~~~\nonumber\\
= \gamma_{1\rightarrow 0}\left(V_M\right)p_1(q,t)-\gamma_{0\rightarrow 1}\left(V_M\right)p_0(q,t),
 \label{pde1}
 \end{eqnarray}
 where $V_M=V(t)-q/C$ is the voltage across the memristor.

The other equation is obtained by replacing
$0\rightarrow 1$ and $1\rightarrow 0$ in Eq. (\ref{pde1}):
\begin{eqnarray}
\frac{\partial p_1(q,t)}{\partial t}+\frac{\partial}{\partial q}\left( \frac{V_M}{R_{1}}p_1(q,t)\right)~~~~~~~~~~~~~\nonumber\\
= \gamma_{0\rightarrow 1}\left(V_M\right)p_0(q,t)-\gamma_{1\rightarrow 0}\left(V_M\right)p_1(q,t).
 \label{pde2}
\end{eqnarray}
The system of Eqs. (\ref{pde1}) and (\ref{pde2}) must be supplemented with
the initial distributions: $p_0(q,t=0)\equiv f(q)$ and
$p_1(q,t=0)\equiv g(q)$.

We note that Eqs. (\ref{pde1}) and (\ref{pde2})  can be considered as generalized continuity equations.
Schematically, the interpretation of various terms in Eq. (\ref{pde1}) is presented in
Fig.~\ref{fig:1}(b). The second term in the left-hand side of Eq.~(\ref{pde1})
describes the flow of probability density through the boundaries of a small charge interval $\Delta q$ (the horizontal arrows in Fig.~\ref{fig:1}(b)).
The right-hand side of Eq.~(\ref{pde1}) represents the flow of probability density between
$p_0(q,t)$ and $p_1(q,t)$ (the vertical arrows in Fig.~\ref{fig:1}(b)).

\subsection{Analytical solutions}  \label{sec:32}

While the fundamental solution of Eqs. (\ref{pde1}) and (\ref{pde2}) delivers
the complete description of circuit behavior, it cannot be found analytically in a closed form for an arbitrary $V(t)$. Thus we confine ourselves to some important particular cases, when such solution can be found.
They are \textit{i}) the limit of small voltages when no transition occurs, and \textit{ii}) unidirectional switching case, when transitions go only from one state to another.

\subsubsection{No switching case}

For those moments of time $t$, capacitor charge $q$, and applied voltage $V(t)$
when no switching practically occurs, Eqs. (\ref{pde1})-(\ref{pde2}) can be simplified to the following independent equations:
\begin{eqnarray}
\frac{\partial p_0(q,t)}{\partial t}+\frac{\partial}{\partial q}\left[ \frac{V_M}{R_{0}}p_0(q,t)\right]=0,
 \label{pde1a}\\
\frac{\partial p_1(q,t)}{\partial t}+\frac{\partial}{\partial q}\left[ \frac{V_M}{R_{1}}p_1(q,t)\right]=0.
 \label{pde2a}
\end{eqnarray}
The general solution of Eqs. (\ref{pde1a})-(\ref{pde2a}) can be found by the method of characteristics~\cite{strauss2007partial} and presented as
\begin{eqnarray}
p_0(q,t)=e^{\frac{t}{CR_{0}}}f\left(qe^{\frac{t}{CR_{0}}}-\int\limits_0^t e^{\frac{\tau}{CR_{0}}}\frac{V(\tau)}{R_{0}} \textnormal{d}\tau \right),\;\;\;\;\;\;
 \label{s1}
\end{eqnarray}
\begin{eqnarray}
p_1(q,t)=e^{\frac{t}{CR_{1}}}g\left(qe^{\frac{t}{CR_{1}}}-\int\limits_0^t e^{\frac{\tau}{CR_{1}}}\frac{V(\tau)}{R_{1}} \textnormal{d}\tau \right),\;\;\;\;\;
 \label{s2}
\end{eqnarray}
where $f(q)$ and $g(q)$ are two arbitrary functions (initial conditions). We reiterate that Eqs. (\ref{s1}) and (\ref{s2}) with $p_{0}(q,0)=f(q)$ and $p_1(q,0)=g(q)$ provide the full solution in the case when the transitions between the states can be neglected.

To illustrate the above solution, we consider the step-like initial probability distribution
\begin{eqnarray}
p_{0}(q,0)&=&\begin{cases}
\frac{1}{q_\beta-q_\alpha} , \mbox{ for } q\in(q_{\alpha},q_{\beta}), \label{eqw21a}\\
0 \;\;\;\; , \;\;\; \mbox{otherwise} , \label{eqw21b}
\end{cases} \\
p_1(q,0)&=&0.
 \label{p10}
\end{eqnarray}
Then in accordance with
Eq. (\ref{s1}) we get the following expression
 for the charge probability distribution at any moment of time:
\begin{eqnarray}
p_{0}(q,t)=\begin{cases}
\frac{e^\frac{t}{CR_{0}}}{q_\beta-q_\alpha} , \mbox{ for } q\in(q_{\alpha}(t),q_{\beta}(t)), \\
0 \;\;\;\; , \;\;\; \mbox{otherwise} ,
\end{cases}\\
 \label{p1t}
\end{eqnarray}
where
\begin{equation}
q_{i}(t)=q_i e^{-\frac{t}{CR_{0}}}+\int_0^t  e^{\frac{\tau-t}{CR_{0}}}\frac{V(\tau)}{R_{0}}\textnormal{d}\tau,
\label{Qt}
\end{equation}
with $i=\alpha, \beta$. Eq. (\ref{Qt}) is similar to the time-dependence of capacitor charge in the classical
RC-circuit subjected to the voltage $V(t)$.

From Eq. (\ref{p1t}) we see that the dissipative nature of memristor-capacitor circuit
leads to the exponential narrowing  of the probability distributions with time.  At long times, $t\gg CR_{0}$, the
distribution approaches Dirac delta function
\begin{equation}
p_0(q,t)=\delta\left(q-\int\limits_0^t  e^{\frac{(\tau-t)}{CR_{0}}}\frac{V(\tau)}{R_{0}}\textnormal{d}\tau\right). \;\;\;
\label{p1tb}
\end{equation}
It is clear Eq. (\ref{p1tb}) is valid for any initial distribution $f(q)$ in the absence of resistance switching events.

Moreover, the initially deterministic state will remain deterministic in the absence of switchings.
For  $q(t=0)=q_0$ and $p_1(q,t=0)=0$,  Eqs.~(\ref{s1}) and (\ref{s2}) lead to
\begin{eqnarray}
p_0(q,t)&=&\delta\left(q-q_0e^{-\frac{t}{CR_0}}-\int\limits_0^t  e^{\frac{(\tau-t)}{CR_{0}}}\frac{V(\tau)}{R_{0}}\textnormal{d}\tau\right), \;
\label{determ} \\
p_1(q,t)&=&0.
\end{eqnarray}
According to Eq.~(\ref{determ}), the evolution of the capacitor charge is
\begin{equation}
q(t)=q_0e^{-\frac{t}{CR_0}}+\int\limits_0^t  e^{\frac{(\tau-t)}{CR_{0}}}\frac{V(\tau)}{R_{0}}\textnormal{d}\tau. \;\;\;
\label{determ2}
\end{equation}

\subsubsection{Unidirectional switching}

Next we consider the region in $q-t$ phase plane,  where
the transitions from $R_{1}$ to $R_{0}$ are forbidden, but the opposite transitions may occur.
In this situation Eqs. (\ref{pde1}) and (\ref{pde2}) can be rewritten as
\begin{eqnarray}
\frac{\partial p_0(q,t)}{\partial t}+\frac{\partial}{\partial q}\left[ \frac{V_M}{R_{0}}p_0(q,t)\right]
&=& -\gamma_{0\rightarrow 1}\left(V_M\right)p_0(q,t), \;\;\;\;\;\;
 \label{pde1b}\\
\frac{\partial p_1(q,t)}{\partial t}+\frac{\partial}{\partial q}\left[ \frac{V_M}{R_{1}}p_1(q,t)\right]
&=& \gamma_{0\rightarrow 1}\left(V_M\right)p_0(q,t). \;\;\;\;\;\;
 \label{pde2b}
\end{eqnarray}
Eqs.~(\ref{pde1b}) and (\ref{pde2b}) are valid when $\gamma_{1\rightarrow 0}\left(V_M\right)=0$.
 The general solution of Eqs. (\ref{pde1b}) and (\ref{pde2b})
can be derived by using the method of characteristics~\cite{strauss2007partial}. As a result
we get
\begin{widetext}
\begin{eqnarray}
p_0(q,t)&=&e^{\frac{t}{CR_{0}}}f\left(qe^{\frac{t}{CR_{0}}}-\int\limits_0^t e^{\frac{\tau}{CR_{0}}}\frac{V(\tau)}{R_{0}} \textnormal{d}\tau  \right)
\exp
\left\{
-\int\limits_0^t  \gamma_{0\rightarrow 1}
\left(
V(\tilde t)-\frac{q}{C}e^{\frac{(t-\tilde t)}{CR_{0}}}
-\frac{e^{\frac{-\tilde t}{CR_{0}}}}{CR_{0}}
\int\limits^{\tilde t}_t  e^{\frac{\tau}{CR_{0}}}V(\tau) \textnormal{d}\tau
\right) \textnormal{d}\tilde t
\right\}, \;\;\; \label{s1a} \\
p_1(q,t)&=&e^{\frac{t}{CR_{1}}}g\left(qe^{\frac{t}{CR_{1}}}-\int_0^t e^{\frac{\tau}{CR_{1}}}\frac{V(\tau)}{R_{1}}\textnormal{d}\tau \right) \nonumber\\
&+& \int\limits_0^t \textnormal{d}\tilde t \gamma_{0\rightarrow 1}
\left(
V(\tilde t)-\frac{q}{C}e^{\frac{(t-\tilde t)}{CR_{1}}}
-\frac{1}{CR_{1}}e^{\frac{-\tilde t}{CR_1}}
\int\limits^{\tilde t}_t \textnormal{d}\tau e^{\frac{\tau}{CR_{1}}}V(\tau)
\right) e^{\frac{(t-\tilde t)}{CR_{1}}}p_0\left( qe^{\frac{(t-\tilde t)}{CR_{1}}}
+\int\limits_t^{\tilde t} \textnormal{d}\tau e^{\frac{(\tau-\tilde t)}{CR_{1}}}\frac{V(\tau)}{R_{1}},
\tilde t \right),\label{s1b}
\end{eqnarray}
\end{widetext}
which are valid in the region $q<CV(t)$, where $\gamma_{1\rightarrow 0}\left(V_M\right)=0$,
and $f(q)$ and $g(q)$ are two arbitrary functions.

\begin{figure}[tb]
\begin{center}
\includegraphics[width=.85\columnwidth]{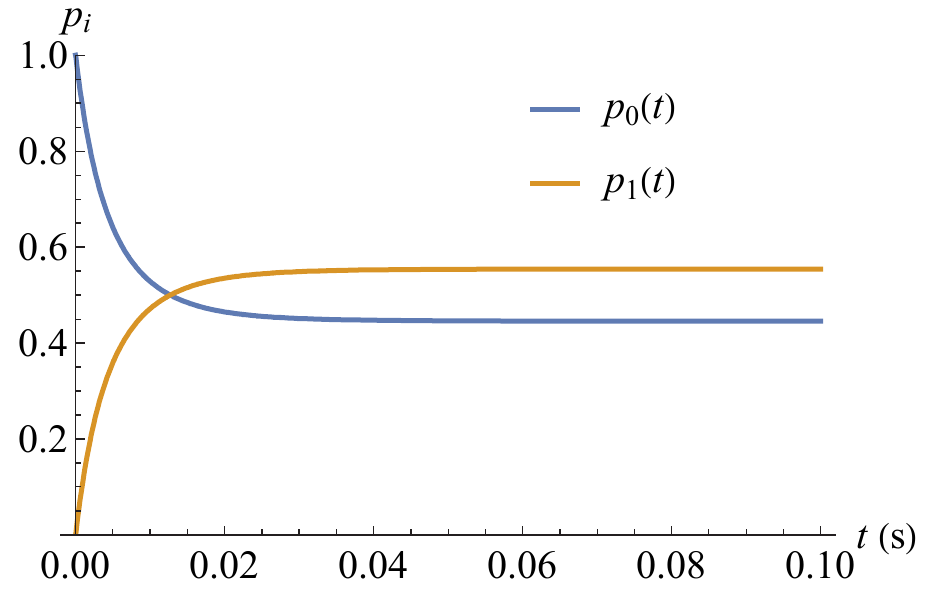}
\caption{\label{fig:2} Probabilities to find the memristive devices in the state 0 and state 1 in the binary memristor-capacitor circuit subjected to a constant voltage $V_a$.
This plot was obtained using Eq.~(\ref{eq:sol2}) with parameter values $C=1$~$\mu$F, $R_0=100$~k$\Omega$, $V_a=0.35$~V, $V_0=0.02$~V, $\tau_0=3\cdot 10^5$~s, $q_0=0$.}
\end{center}
\end{figure}

By the interchanging indexes
$0\rightarrow 1$ and $1\rightarrow 0$ in Eqs. (\ref{s1a}) and
 (\ref{s1b}) we can also find the solutions of  Eqs. (\ref{pde1}) and (\ref{pde2})
 for the region $q>CV(t)$, where $\gamma_{0\rightarrow 1}\left(V_M\right)=0$.
Note that if there are no transitions at all ($\gamma_{0\rightarrow 1}\equiv 0$), then the results (\ref{s1a}) and (\ref{s1b}) coincide with Eqs.~(\ref{s1}) and (\ref{s2}).

As an example of the theory above let us consider the case of deterministic initial conditions and constant applied voltage: $q(t=0)=q_0$, $p_1(q,t=0)=0$, and $V(t)=V_a$. Then, Eq.~(\ref{s1a}) simplifies to
\begin{equation}\label{eq:sol1}
  p_0(q,t)=\delta\left( q-q(t)\right)e^{-\frac{CR_0}{\tau_0}\left[\textnormal{Ei}\left(\frac{V_a-\frac{q}{C}}{V_0}e^{\frac{t}{CR_0}}\right)-\textnormal{Ei}\left(\frac{V_a-\frac{q}{C}}{V_0}\right)\right]},
\end{equation}
where $\textnormal{Ei}(x)$ is the exponential integral function and
\begin{equation}\label{eq:sol1a}
  q(t)=q_0e^{-\frac{t}{CR_0}}+V_a C\left( 1-e^{-\frac{t}{CR_0}}\right).
\end{equation}
The integral of Eq.~(\ref{eq:sol1}) over $q$ from minus to plus infinity gives the probability to find the memristor in the state 0 at time $t$:
\begin{equation}\label{eq:sol2}
  p_0(t)=e^{-\frac{CR_0}{\tau_0}\left[\textnormal{Ei}\left(\frac{V_a-\frac{q_0}{C}}{V_0}\right)-\textnormal{Ei}\left(\frac{V_a-\frac{q_0}{C}}{V_0}e^{-\frac{t}{CR_0}}\right)\right]},
\end{equation}
where $q(t)$ is defined by Eq.~(\ref{eq:sol1a}). We note that $p_1(t)$ can be found from $p_0(t)+p_1(t)=1$.

Fig.~\ref{fig:2} shows  the evolution of $p_0(t)$ and $p_1(t)=1-p_0(t)$ found with the help of Eq.~(\ref{eq:sol2}). While it may look that the memristor switching is incomplete on average, in fact, the circuit dynamics is a two-time-scale process where the fast initial relaxation (such as the visible one in Fig.~\ref{fig:2}) transforms into a very slow one occurring on the time scale of $\tau_0$ (as a consequence of Eq.~(\ref{eq:gamma01})).
Clearly, the fast initial relaxation decelerates with time as the voltage builds up across the capacitor.

\begin{figure}[t]
\begin{center}
\includegraphics[width=.85\columnwidth]{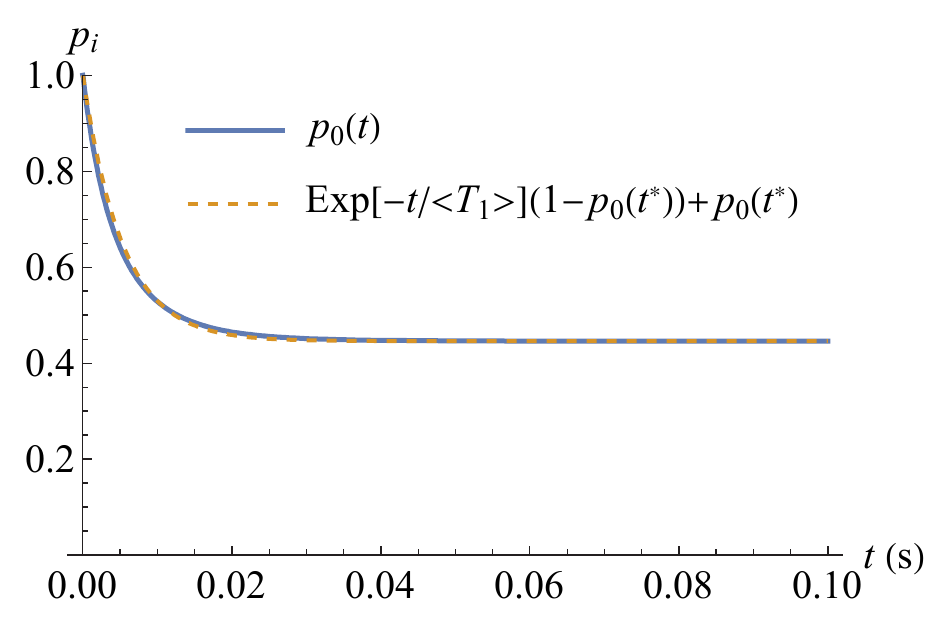}
\caption{\label{fig:3} The probability to find the memristive device in state 0, $p_0(t)$  is well approximated by an exponential decay curve based on $ \left< T_1 \right>$ (Eq.~(\ref{eq:swtime})). The exponential decay curve was obtained using $\left< T_1 \right>=5.3$~ms and $t^*=1$~s. The probability $p_0(t)$ curve is the same as in Fig.~\ref{fig:2}.}
\end{center}
\end{figure}

 The mean switching time for the dynamics in Fig.~\ref{fig:2} can be calculated using
\begin{equation}\label{eq:swtime}
  \left< T_1 \right>=\frac{1}{p_1(t^*)}\int\limits_0^{t^*} t \frac{\textnormal{d}p_1(t)}{\textnormal{d}t}\textnormal{d}t,
\end{equation}
where $t^*$ is the  characteristic saturation time for the  fast initial dynamics of $p_i(t)$. Taking $t^*=1$~s, one can find  that $\left< T_1 \right>=5.3$~ms, which is in perfect agreement with the behavior in Fig.~\ref{fig:2}. The exponential relaxation with the mean switching time $ \left< T_1 \right>$ provides an excellent approximation for the initial interval of fast evolution of $p_0(t)$, see Fig.~\ref{fig:3}. On long times, the  relaxation can be approximated by $p_0(t)\simeq p_0(t^*)\textnormal{exp}(-t/\tau_0)$.

Moreover, by using Eq.~(\ref{eq:sol2}) one can find $p_0(t^*)$ in the most interesting parameter region $(V_a-q_0/C)\gg V_0$ corresponding to the initial interval of fast evolution.
This is accomplished employing the asymptotic expansion of the exponential integral function
$\textnormal{Ei}(x)=e^x/(x-1)\left(1+\textnormal{O}(1/x^2)\right)$, when $x\rightarrow+\infty$. For the time interval $\left< T_1 \right>\lesssim t\ll \tau_0$, we can omit the second term in the square brackets in Eq.~(\ref{eq:sol2}) and use  the asymptotic expansion for the first one. This leads us to the following expression for the probability of no switching  event
\begin{equation}\label{eq:sol3}
  p_0(t^*)=
  \exp
  \left[
  -\frac{CR_0}{\tau_0}
  \frac{e^
  {\left(
  \frac{V_a-q_0/C}{V_0}
  \right)}}{\frac{V_a-q_0/C}{V_0}-1}
  \right].
  \end{equation}
For the same parameter values as in Fig.~\ref{fig:2}, Eq.~(\ref{eq:sol3}) gives $p_0(t^*)\approx0.447$, which is in an excellent agreement with the result  $p_0(t^*)=0.446$ obtained numerically from Eq.~(\ref{eq:sol2}) .

\section{Discission} \label{sec:4}

The method used to model the memristor-capacitor circuit above can be straightforwardly extended to other circuits.
Consider, for instance, a circuit composed of $N$ $G$-state stochastic memristors, $K$ capacitors and $J$ inductors. In the general case,
the simulation of such a circuit requires $G^N$ probability distribution functions of $K+J$ variables and time. Circuits with symmetries may require less functions to represent their states (see Ref.~\onlinecite{dowling2020probabilistic} for examples). The number of independent reactive variables can be less than $K+J$. For example, if the external voltage $\mathcal{E}(t)$ is applied directly across some capacitor, then its charge is not an independent variable.

It is anticipated that the general evolution equation can be written similarly to Eqs. (\ref{pde1}) and (\ref{pde2}). Introducing the sets of capacitive and inductive variables, $Q$ and $I$, the evolution equation for a particular state $i$ is formulated as
\begin{widetext}
\begin{equation}
\frac{\partial p_i(Q,I,t)}{\partial t}+\sum_j\frac{\partial}{\partial Q_j}\left[ \dot{Q}_jp_i(Q,I,t)\right]+\sum_j\frac{\partial}{\partial I_j}\left[ \dot{I}_jp_i(Q,I,t)\right]
=\sum_{j\neq i}\left[\gamma_{ji}p_j(Q,I,t)-\gamma_{ij}p_i(Q,I,t)\right],~~~~~~~
 \label{pdemn}
\end{equation}
\end{widetext}
where the memristive switching rates corresponds to the flip of a single memristor and depend on the voltage across the same memristor in a particular circuit configuration. In
order to close Eq.  (\ref{pdemn}) we need to express the full derivatives
$\dot{Q}_j$ and $\dot{I}_j$ as functions of $R, Q, I$ by using the Kirchhoff's circuit laws.
 If there are no transitions, then the RHS of  Eq.  (\ref{pdemn})
turns to zero and this equation coincides with the continuity equation for the probability
density $p_i(Q,I,t)$ as it should be.



In conclusion, we have introduced a powerful analytical approach to model heterogeneous stochastic circuits. A simple example was considered in detail and the recipe to apply the approach to other circuits has been formulated. Compared to the traditional Monte Carlo simulations, the proposed approach can be used to derive analytical expressions describing the circuit dynamics on average.

The data that support the findings of this study are available from the corresponding author upon reasonable request.



%
%

%


\bibliographystyle{apsrev4-1}
\bibliography{memcapacitor}

\end{document}